\documentclass[11pt,a4paper]{article}
\usepackage{bm,jcappub,Macro,tabularx}
\usepackage{url}
\usepackage{hyperref}
\usepackage{url}
\usepackage{epstopdf}
\usepackage{enumitem}
\usepackage{colortbl}
\usepackage{color}
\def\la{\langle}
\def\ra{\rangle}
\def\n{\noindent}
\def\be{\begin{equation}}
\def\ee{\end{equation}}
\def\ben{\begin{eqnarray}}
\def\een{\end{eqnarray}}
\def\nn{\nonumber}

\def\bk{{\bf k}}

\def\bk{{\bf k}}

\def\bx{{\bf x}}
\def\2p{{(2\pi)^2}}

\def\be{\begin{equation}}
\def\ee{\end{equation}}
\def\beq{\begin{equation}}
\def\eeq{\end{equation}}
\def\ben{\begin{eqnarray}}
\def\een{\end{eqnarray}}

\def\nn{{\nonumber}}

\newcommand{\beqa}{\begin{eqnarray}}
\newcommand{\eeqa}{\end{eqnarray}}

\def\ikap0{{\cal J}_{\theta_0}(r)}

\def\one1{\langle \kappa_{(i)}\kappa_{(j)} \rangle}
\def\one{{[\bar \xi^{(ij)}]}}

\def\ba{\begin{eqnarray}}
\def\ea{\end{eqnarray}}

\def\bk{{\bf k}}
\def\bq{{\bf q}}

\def\pfact{{P_{\delta}(k)P_{\delta}(q_b)}}

\title{The Integrated Bispectrum in Modified Gravity Theories}
\author{Dipak Munshi}
\affiliation{Astronomy Centre, School of Mathematical and Physical Sciences,\\ University of Sussex, Brighton BN1 9QH, U.K.}
\emailAdd{D.Munshi@sussex.ac.uk}
\abstract{Gravity-induced non-Gaussianity can provide important clues to Modified Gravity (MG) Theories.
Several recent studies have suggested using the {\it Integrated Bispectrum} (IB) 
as a probe for squeezed configuration of bispectrum. 
Extending previous studies on the IB,
we include redshift-space distortions to study a class of (parametrised) MG
theories that include the string-inspired Dvali, Gabadadze \& Porrati (DGP) model.
Various contributions from redshift-space distortions are derived in a transparent
manner, and squeezed contributions from these terms are derived separately.
Results are obtained using the Zel'dovich Approximation (ZA).
Results are also presented for projected surveys (2D).
We use the Press-Schechter (PS) and Sheth-Torman (ST) mass functions to compute the 
IB for collapsed objects that can readily be extended to peak-theory based approaches.
The {\em cumulant correlators} (CCs) generalise the ordinary {\em cumulants} and are known to 
probe collapsed configurations of higher order correlation functions.
We generalise the concept of CCs to halos of different masses.
We also introduce a generating function based approach to analyse more
general non-local biasing models. The Fourier representations of 
the CCs, the skew-spectrum, or the kurt-spctra are discussed in this context.
The results are relevant for the study of the Minkowski Functionals (MF) of collapsed tracers in redshift-space.}
\begin{document} 
\maketitle
%
\section{Introduction}
\label{sec:intro}
Ongoing and recently completed CMB experiments Planck\footnote{Planck: \href{http://www.cosmos.esa.int/web/planck/}{\tt  http://www.cosmos.esa.int/web/planck/}},
ACT\footnote{ACT: \href{http://www.physics.princeton.edu/act/}{\tt http://www.physics.princeton.edu/act/}},
SPT\footnote{SPT: \href{http://pole.uchicago.edu/}{\tt http://pole.uchicago.edu/}}.
have provided us with a standard cosmological model. However, many
questions regarding the nature of the dark matter or dark energy or,
equivalently, possible modification of General Relativity (GR)
remain unanswered.

While, a finely tuned cosmological constant can explain the accelerated expansion of the Universe \citep{Perlmutter,Riess},
modifying the laws of gravity remains another possibility \citep{Clifton12,Joyce15}. The rules of gravity are
highly constrained at solar system \citep{OS03}, as well as, at astrophysical scales \citep{Vikram13}. However 
at cosmological scales modifications of gravity are poorly constrained. 
The background cosmological evolution in MG theories can be difficult to distinguish from that in
many dark energy scenarios. Evolution of perturbations at low redshift can however provide a useful
diagnostic to differentiate between these two scenarios.  

Large-scale structure (LSS) surveys constrain cosmology with ever higher precision 
(e.g. \citep{Anderson} and references therein).
Over the past decade or so there has been a major progress
in mapping the galaxy distribution using spectroscopic as well
as photometric redshifts (BOSS\footnote{Baryon Oscillator Spectroscopic Survey: \href{http://www.sdss3.org/surveys/boss.php}{\tt http://www.sdss3.org/surveys/boss.php}}
\citep{EW}
Wiggle\footnote{Dark Energy Survey :  \href{http://wigglez.swin.edu.au/}{\tt http://wigglez.swin.edu.au/}}
\citep{DJA}
DES\footnote{Dark Energy Survey: \href{http://www.darkenergysurvey.org/}{\tt http://www.darkenergysurvey.org/}}
\citep{DES}
EUCLID\footnote{EUCLID: \href{http://www.euclid-ec.org/}{\tt http://www.euclid-ec.org/}}
\citep{LAA}).
The resulting maps and their analysis have already revolutionised cosmology by putting the most stringent
constraints on the growth of structures as well as expansion history of the Universe.

Traditionally, the two-point correlation function in real space, or, equivalently,
the power spectrum, has always been used to analyse the galaxy clustering \citep{galaxy_red}. 
Modelling the evolution of the power spectrum in MG theories have done using many different extensions of
perturbative techniques at large scales and halo model prescriptions at small scale \citep{BraxValageas12,BraxValageas13}.
Techniques based on resummation methods \citep{crocce} coupled to an {\em eikonal} approximation have also been developed \citep{Atsushi}.
Numerical simulations are often  used to test and validate such approaches \citep{Oyaizu,LiZhao, Winther}.

The problem of estimation of the power spectrum from galaxy surveys has been dealt with using a flat-sky approach as well as using an all-sky formalism.
While most initial estimations were performed in projection (2D)\citep{SzSz,Bernard_angular} and an inversion technique was subsequently employed to
reconstruct the 3D power spectrum \citep{BaughEfstathiou93}; with the improvement in data quality, the recent spectroscopic surveys
allow data to be analysed in three dimensions (3D) \citep{HeavensTaylor,Cyril16}. However, measured galaxy power spectrum
is a biased tracer of the underlying matter power-spectrum
and the degeneracy between the parameters describing growth, normalisation and bias (denoted as $f,\sigma_8, b$) can not be broken by power spectrum alone.
 
Gravity-induced higher order correlation functions or their Fourier representations, the
so called higher order polyspectra, can provide important clues to structure formation scenarios
by breaking the above-mentioned degeneracy (see Ref.\cite{review} for a review) thus motivating a flurry of
recent activity in developing estimators for secondary non-Gaussianity.
Due to the rapid progress in large scale surveys, it is now possible to estimate 
the higher order correlation functions \citep{GilMarin}.
Analytical modelling of evolution of bispectrum in MG theories is more difficult \citep{BraxValageas13}.
Modelling clustering of galaxies involves several different but related
steps including modelling the dark matter clustering \citep{crocce,okamura,
CarlsonWhite,Pietroni,ValageasNishimichi}
using e.g. semi-analytic prescription.
To relate the clustering of galaxies to clustering
of underlying dark matter distribution typically a well motivated
biasing scheme that can reproduce the simulation results
\citep{NishimichiTaruya,McDonaldRoy,SaitoBaldauf,Biagetti}. In our study we will
use the bias computed using the halo model based approach of Ref.\citep{MoJingWhite}.
We also consider the approach described in Ref.\citep{McDonaldRoy}.
Many recent studies have recently focused on how to relate the real space clustering and  
the observed redshift-space distribution of galaxies \citep{VlahSeljak,TaruyaNishimici,ReidWhite}.
We will consider the impact of modifying gravity on the {\em squeezed} configuration of galaxy bispectrum in redshift-space.

The bispectrum represents the lowest order in non-Gaussinaity and
is defined as a function of three wave-vectors that describes a triangular configuration in the harmonic domain.
Estimation of the bispectrum for each possible triangular configuration from data is far more demanding than the estimation of power spectrum
in the presence of a complicated survey geometry and anisotropic noise \citep{Komatsu_review}.
This has motivated the development of estimators which are sensitive to {\em collapsed} configurations of the bispectrum. 
The cumulant correlators (CCs) are a natural generalisation of the one-point cumulants. They represent higher order correlation functions
collapsed to two-points and provide an alternative route to the study of higher order correlation hierarchy
and are well studied in the literature in the perturbative regime \cite{francis}, as well as, in
the highly nonlinear regime using the {\em hierarchical ansatz} (HA) \cite{gen}. 
The Fourier representation of {\em optimised}
lower order CCs i.e. the skew-spectrum (third order)\cite{skewspec} and
kurt-spectrum (fourth order) \cite{kurtspec}
were shown as an important form of data compression in 2D as well as in 3D.

In recent years it has been realised that the
measurements of the power spectrum in a sub-volume of the survey is statistically correlated to
the average density contrast in that subvolume. This correlation of the {\em position-dependent}
power spectrum and the average density contrast was recently used
to define an estimator for the bispectrum (see Ref.\cite{komatsu2} for a complete list of references).
Such an estimator is sensitive to the squeezed configuration of the bispectrum.
Mathematical structure of the CCs and their Fourier representation is
related to the position-dependent power spectrum was underlined  recently in Ref.\citep{MunshiColes}.
The primary aim of this paper is to extend the concept of IB to redshift-space  for a class of MG theories,
as well as in projection (2D). Generalising Ref.\citep{MoJingWhite} we also derive the CCs for collapsed objects.
We also employ the generating function formalism to relate the CCs of collapsed objects
to that of underlying mass distribution. 

This article is arranged as follows. In \textsection\ref{sec:cc}
we introduce the CCs for the collapsed
objects of halo model. We also show how the CCs of high density peaks
can also be computed using this approach.
In \textsection\ref{sec:IntBiTri}
we recapitulate the recently introduced concept of integrated bispectrum
and trispectrum and compute these quantities for collapsed objects.
Different models of non-local bias are considered in \textsection\ref{sec:nonlocal}.
The effect of redshift space distortion is analysed in \textsection\ref{sec:red}.
The specific $\gamma$ models are briefly introduced in \textsection\ref{sec:MG}.
The integrated bispectrum is introduced in \textsection\ref{sec:red_IB}.
The projected integrated bispectrum is presented in \textsection\ref{sec:2D}.
Finally in \textsection\ref{sec:conclu} we summarize the results and present
our conclusions.
%
%
\section{Cumulant Correlators of Collapsed Objects}
\label{sec:cc}
We will compute the cumulants and CCs
for the halos following Ref.\cite{MoWhite} and  for peak Ref.\citep{peak}
and relate them to the underlying cumulans of density distribution.
The density contrast of halos at a comoving coordinate ${\bf x}$ will be denoted by $\delta^{(h)}({\bf x},a)$, and is related to 
the underlying density contrast of matter distribution $\delta({\bf x},a)$
through a generic local, nonlinear and deterministic bias 
$\delta^{(h)}({{\bf x},a}) = b[\delta({\bf x},a)]$. The results
will be extended to the case of non-local bias in \textsection\ref{sec:nonlocal}; 
inclusion of {\em stochastic} bias can also be done in a straight-forward manner.
We will assume that the halo overdensity can be expanded in a Taylor series in terms of the underlying density contrast $\delta$: 
\ben
\delta^{(h)}({\bf x}) \equiv b[\delta(\bx)] = \sum_k\, {b_k \over k!}\, [\delta({\bf x})]^k.
\label{eq:halo_bias}
\een
The cumulants ${\mathbb S}_n$ and CCs ${\mathbb C}_{pq}$ for the underlying dark matter distribution are defined in terms of the following expressions:
\ben
&& \la [\delta({\bf x},a)]^p \ra_c = {\mathbb S}_n \la[\delta({\bf x},a)]^{2}\ra_c^{p-1};\\
&& \la [\delta({\bf x}_1,a)]^p[\delta({\bf x}_2,a)]^q\ra_c = {\mathbb C}_{pq} \la[\delta({\bf x},a )]^2 \ra_c^{p+q-2}
\la\delta({\bf x}_1,a)\delta({\bf x}_2,a) \ra_c.
\een
The corresponding cumulants ${\mathbb S}^{(h)}_n$ and CCs ${\mathbb C}^{(h)}_{pq}$ for halos are defined using similar expressions:
\ben
&& \la [\delta^{(h)}({\bf x},a)]^p \ra_c = {\mathbb S}^{(h)}_n \la[\delta^{(h)}({\bf x},a)]^{2}\ra_c^{p-1};\\
&& \la [\delta^{(h)}({\bf x}_1,a)]^p[\delta^{(h)}({\bf x}_2,a)]^q\ra_c = {\mathbb C}^{(h)}_{pq} \la[\delta^{(h)}({\bf x},a )]^2 \ra_c^{p+q-2}
\la\delta^{(h)}({\bf x}_1,a)\delta^{(h)}({\bf x}_2,a) \ra_c.
\een

\begin{figure}
\vspace{1.25cm}
\begin{center}
{\epsfxsize=13. cm \epsfysize=4.7 cm 
{\epsfbox[32 425 547 585]{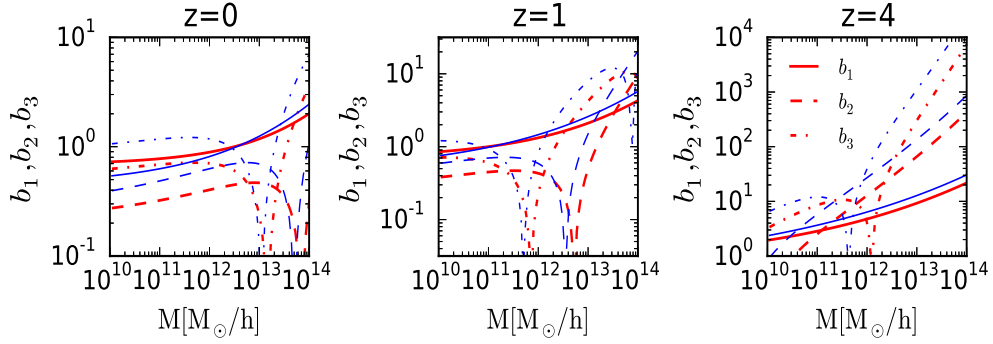}}}
\caption{The first three bias parameters $b_1,b_2$ and $b_3$ are shown
as a function of halo mass. The bias parameters are defined in
Eq.(\ref{eq:b1})-Eq.(\ref{eq:b3}).
We have used a ST (thick lines) and PS (thin lines) mass function defined in Eq(\ref{eq:ST}).
The plots from left to right correspond respectively to $z=0$, $z=1$ and $z=4$.}
\label{fig:b1b2b3}
\end{center}
\end{figure}

Our aim here is to express the cumulants and CCs of halos in terms of their underlying dark matter counterparts. 
We will use the generating function formalism developed in Ref.\citep{Ber92} to 
construct the CCs for the biased tracers (halos or peaks) and express them in terms of the same 
for underlying mass distribution. The results are relevant for perturbative regime.
The n-th order of perturbative expansion of an arbitrary field $F(\bx,a)$ defined as
$F^{(n)}(\bx,a)$ with respect to $\delta(\bx,a)$ is defined as follows \citep{Ber92}:
\ben
\la {F}^{(n)}\ra_c = 
{{\int \la {F}^{(n)}({\bf x}, a)\delta^{(1)}({\bf x}_1,a)\cdots \delta^{(1)}({\bf x}_n,a)\ra_c\; d^3{\bf x}\,d^3{\bf x}_1 \cdots d^3{\bf x}_n}
\over 
({\int \la\delta^{(1)}({\bf x},a)\,\delta^{(1)}({\bf x^{\prime}},a) \ra 
d^3{\bf x}\, d^3{\bf x}^{\prime}})^n}.
\een
Here $\delta^{(1)}({\bf x},a)$ is the
linear approximation for $\delta({\bf x},a)$, and only connected diagrams are taken into
account, which explains the subscript $c$. Throughout we will assume that the
initial density field $\delta$ is Gaussian, though it is possible to incorporate non-Gaussian initial condition.

For two arbitrary fields $A(\bx)$ and $B(\bx)$, we have the following properties 
for the generating functions \citep{Ber92}:
\ben
\label{eq:gen_1}
&& {\cal G}_{A+B}(\tau) = {\cal G}_A(\tau)+{\cal G}_B(\tau); \quad
{\cal G}_{AB}(\tau) = {\cal G}_A(\tau){\cal G}_B(\tau); \quad\\
&&{\cal G}_{\partial_i A \; \partial_i B}(\tau) = 0; \quad
 {\cal G}_{\partial_i\partial_j A \;\partial_j\partial_i B}(\tau) = {1 \over 3}{\cal G}_{\triangle A\triangle B}(\tau).
\label{eq:gen_n}
\een
The generating function ${\cal G}_{F}(\tau)$ for the vertices for any random field $F({\bf x},a)$ is given by:
\ben
{\cal G}_{F}(\tau) = \sum^{\infty}_{n=1} {\la {F}^{(n)} \ra_c \over n!}\tau^n.
\een
We will denote the generating function of underlying density by ${\cal G}_{\delta}(\tau) = \sum^{\infty}_{n=1}\, [{\nu_n/n!}]\, \tau^n$; \;
$\nu_n=\la \delta^{(n)}\ra_c$. Similarly, we will denote the 
corresponding function for the divergence of velocity $\theta$ as: ${\cal G}_{\theta}(\tau) = \sum^{\infty}_{n=1}\, [{\mu_n/n!}]\, \tau^n$
where $\mu_n=\la \theta^{(n)}\ra_c$. Similar expressions for halos will be defined below and we will denote them with the superscript $^{(h)}$.
Using the generating function formalism of Ref.\cite{Ber92} which ensures
that ${\cal G}_{\delta^q}(\tau)= [{\cal G}_{\delta}(\tau)]^{q}$ we can prove 
${\cal G}^{(h)}(\tau) = b[{\cal G}_{\delta}(\tau)]$ as follows:
\ben
{\cal G}^{(h)}(\tau) = \sum^{\infty}_{n=1}\, {\nu^{(h)}_n \over n!}\, \tau^n
= \sum^{\infty}_{k} {b_k \over k!} [{\cal G}(\tau)]^k = b[{\cal G}(\tau)].
\label{eq:generating}
\een 
Using Eq.(\ref{eq:generating}) the first few halo vertices $\nu_k^{(h)}$ can be expressed in terms of the underlying vertices $\nu_k$ and
the parameters $b_k$ as follows:
\ben
\label{eq:cc1}
&& \nu_1^{(h)} = b_1;\\
&& \nu_2^{(h)} = (b_2 + b_1 \nu_2); \\
&& \nu_3^{(h)} = (b_3 + 3 b_2 \nu_2 + b_1 \nu_3); \\
&& \nu_4^{(h)} =  (b_4 + 6 b_3 \nu_2 + 3 b_2 \nu_2^2 + 4 b_2 \nu_3 + b_1 \nu_4).
\label{eq:cc4}
\een
At this stage the bias parameters $b_k$ are left completely arbitrary.
However, we will use the PS (ST) mass functions to compute these parameters.
Results will be also valid for peaks where the bias functions can be
replaced with the bias functions for peaks beyond a threshold.

The cumulants ${\mathbb S}_n^{(h)}$ are defined in terms of the vertices $\nu^{(h)}_n$ \citep{Fry84,Ber92}:
\ben
\label{eq:S_N}
&& {\mathbb S}^{(h)}_3= 3\nu^{(h)}_2; \\
&& {\mathbb S}^{(h)}_4 = 4\nu^{(h)}_3 + 12[\nu^{(h)}_2]^2; \\
&& {\mathbb S}^{(h)}_5 = 5\nu^{(h)}_4 + 60[\nu_2^{(h)}][\nu_3^{(h)}] + 60 [\nu^{(h)}_2]^3.
\label{eq:S_Np}
\een
In the perturbative regime following relations hold \cite{Ber92,Ber94}:
\ben
\label{eq:S3}
&& {\mathbb S}_3 = {34\over 7} + \gamma_1; \\
&& {\mathbb S}_4 = {60712 \over 1323} + {62 \over 3}\gamma_1 + {7 \over 3}\gamma_1^2 +{2 \over 3}\gamma_2.
\label{eq:S4}
\een
\begin{figure}
\vspace{1.25cm}
\begin{center}
{\epsfxsize=13. cm \epsfysize=4.7 cm 
{\epsfbox[32 425 547 585]{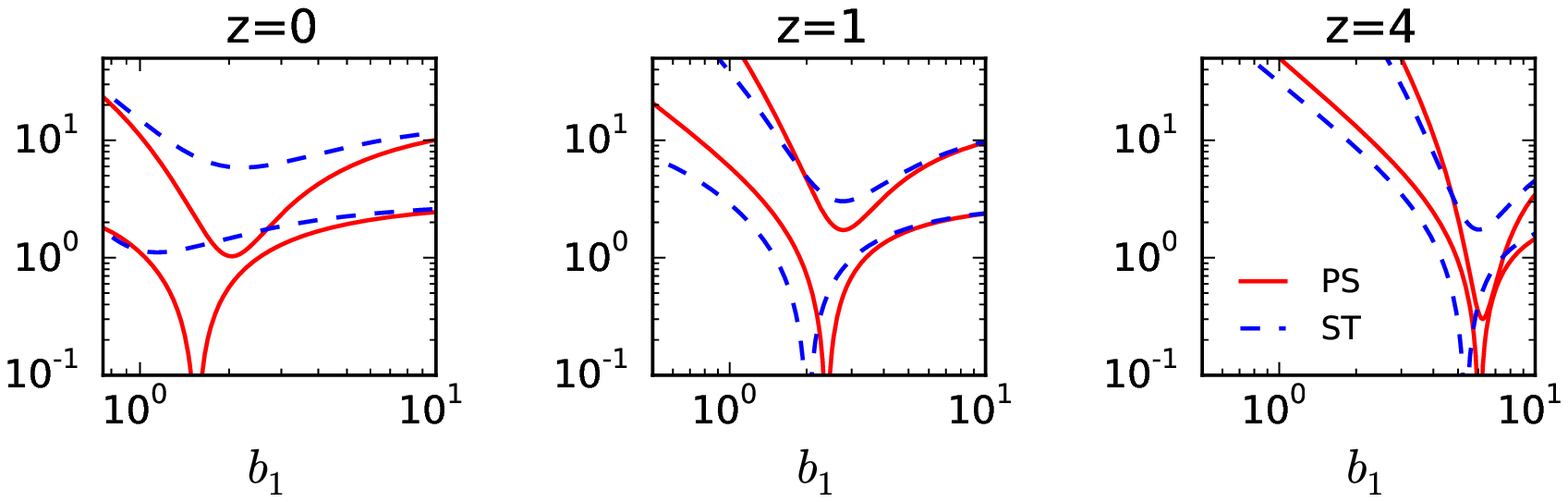}}}
\caption{We plot the normalised cumulants ${\mathbb S}^{(h)}_3$ and ${\mathbb S}^{(h)}_4$
 defined respectively
in Eq.(\ref{eq:H3}) and Eq.(\ref{eq:H4}) for halos as function of the bias of the halos $b_1$ (defined in Eq.(\ref{eq:b1})).
From left to right the panels correspond respectively to redshifts $z=0$, $z=1$ and $z=4$.
The solid (dashed) lines correspond to PS (HT) formalism. The halo cumulants ${\mathbb S}^{(h)}_n$
depend on underlying density cumulants ${\mathbb S}_n$ Eq.(\ref{eq:S3})-Eq.(\ref{eq:S4}). We have assumed a locally
power-law power law with $n_{\rm eff}=-1.5$ to evaluate the parameter $\gamma_1$.}
\label{fig:S3S4}
\end{center}
\end{figure}
The CCs take the following form:
\ben
\label{eq:c21_DM}
&& {\mathbb C}^{(h)}_{21} = 2\nu^{(h)}_2; \\
&& {\mathbb C}^{(h)}_{31} = 3\nu^{(h)}_3 + 6\nu^{(h)}_2; \\
&& {\mathbb C}^{(h)}_{41} = 4\nu^{(h)}_4+36[\nu^{(h)}_2][\nu^{(h)}_3]+24[\nu^{(h)}_2]^3. 
\label{eq:c51}
\een
The CCs satisfy a factorisation property in the large-separation limit [$\xi_{12}(|{\bf x}_1 -{\bf x}_2|) < \sigma^2(R_0)$]:
${\mathbb C}^{(h)}_{pq} = {\mathbb C}^{(h)}_{p1}{\mathbb C}^{(h)}_{q1}$. Here, $\sigma^2(R_0)$ is the variance of the
smoothed density field, and $\xi_2(|{\bf x}_1 -{\bf x}_2|$ represents the two-point correlation function.
A tophat smoothing window with a radius $R_0$ is assumed (to be introduced later in Eq.(\ref{eq:tophat})).
In the quasilinear regime with a tophat smoothing window the CCs have the following expressions \citep{francis}: 
\ben
\label{eq:c21}
&& {\mathbb C}_{21} = {68\over 21}+ {\gamma_1 \over 3}; \\
&& {\mathbb C}_{31} = {11710 \over 441} + {61 \over 7}\gamma_1 + {2 \over 3}\gamma_1^3 +{\gamma_2\over 3}.\\
&& {\mathbb C}_{41} = 353.1 + 205.1\gamma_1 + 38.67\gamma_1^2 +2.33\gamma_1^3 +12.03\gamma_2
+ 2.22\gamma_1\gamma_2 -0.22\gamma_3
\label{eq:c41}
\een
where $\gamma_p = [{d^p \log\,\sigma^2(R_0)/d(\log R_0)^p}]$.
For a power-law power spectrum $P(k)\propto k^{n_{\rm eff}}$, we have $\gamma_1=-(n_{\rm eff}+3)$
and $\gamma_p=0$ for $p>1$. The perturbative results are valid at scales where $\sigma^2(R_0) \le 1$.
Results based on hierarchical ansatz (HA) have been used in highly nonlinear regime with limited success \citep{BS}.
The generating function ${\cal G}_{\delta}(\tau)$ in the quasilinear regime can be computed analytically using the Euler-Continuity-Poisson system. 
The effect of smoothing can also be incorporated analytically in case of tophat window.
For diagrammatic technique based results on computation of CCs using HA see Ref.\citep{Melott1}.
In recent years there has been progress in deriving the cumulants using the  {\em large-deviation} statistics Ref.\citep{Uhelmann}.

Beyond the perturbative regime, due to its highly nonlinear nature there is no analytical description of gravitational clustering. 
The halo model based semi-analytical approaches are remarkably successful in predicting the 
distribution of mases and spatial clustering of halos and the underlying dark matter distribution \citep{halo}. 
Two variants of halo models are more popular than others. 
The Sheth-Torman (ST) and Press-Schechter (PS) mass functions can both be expressed using the following parametrization 
(see Ref.\citep{halo} for a review):
\ben
\label{eq:ST}
&& \nu f(\nu) = A(p)[1+ (q\nu)^{-p}] \, {q\nu \over 2\pi} \, \exp\left (-q\nu/2 \right ).  
\een
The normalisation $A(p)$ and $\nu$ are defined as follows:
\ben
&& A(p) = \left [1+ 2^{-p}\Gamma(1/2-p)/\sqrt{p} \right ]^{-1};\quad  
\nu = {\delta^2_{\rm sc}(z)\over \sigma_{L}^2(m)}; \quad {\delta_{\rm sc}(z)\over (1+z)} \equiv {3 \over 5}{3\pi \over 2}^{2/3}.
\een
Here, $\sigma_L^2(m)$ is the variance in the initial density fluctuation field when smoothed with a 
tophat filter of scale $R_0 = (3m/4\pi\bar\rho)^{1/3}$ extrapolated to the present epoch using linear theory;
$\bar \rho$ represents the comoving density of the background. For a smoothing radius $R_0$ the variance  $\sigma^2_{L}(m)$ 
is computed from the linear power spectrum  $P_{L}(k)$ using a tophat window $W_{\rm TH}$:
\ben
&& \sigma^2_{L}(m) \equiv \int {dk \over k} {k^3 P_{L}(k) \over {2\pi^2}} |W_{\rm TH}(kR_0)|^2; \quad
W_{\rm TH}(x) = {3\over x^3}\left [ \sin(x) -x\cos(x)\right ].
\label{eq:tophat}
\een
For the PS mass function 
we have $p=0$ and $q=1$. The ST function corresponds to $p=0.3$ and $q=0.75$ (See Ref.\citep{Warren,Matsubara} 
for ``$W_+$'' and ``Mice'' mass functions).

In all halo models the clustering of halos are described using bias parameters $b_k(m,z)$ that
are functions of redshift of formation $z_1$ and the mass of the halo $m$.
The halo bias parameters depend on parameters $a_k$, which depend on the dynamics of the spherical
collapse, and parameters  $E_k^{\prime}= E_k+e_k$ (to be defined below) as follows:
\ben
\label{eq:b1}
&& b_1(m,z_1) = 1+ E_1^{\prime}; \\
&& b_2(m,z_1) = 2(1+a_2)E_1^{\prime}+ E_2^{\prime};
\label{eq:b2}\\
&& b_3(m,z_1) = 6(a_2+a_3)E_1^{\prime} +3(1+2a_2)E_2^{\prime} + E_3^{\prime}.
\label{eq:b3}
\een
The $a_k$ parameters are given by the following series expansion in the density contrast $\delta$ 
which describes the spherical collapse as a function of refshift $z$
starting from an initial density contrast $\delta_0$ \citep{Ber94}: 
\ben
{\delta_0 \over 1+z} = \sum^{\infty}_{k=0} a_k \delta^k = 
\delta -{17 \over 21}\delta^2 + {341 \over 567}\delta^3 - {55805 \over 130977}\delta^4 +\cdots
\een
The $e_k$ parameters are defined as follows \cite{MoJingWhite}:
\ben
&& e_1 = {q\nu-1 \over \delta_{sc}(z_1)};  \quad
e_2 = {q\nu\over \delta_{sc}(z_1)} \left({q\nu -3 \over \delta_{sc}(z_1)} \right )\quad
e_3 = {q\nu \over \delta_{sc}(z_1)} \left ( {q\nu-3 \over \delta_{sc}(z_1)} \right )^2.
\een
The first few $E_k$ parameters are listed below \citep{MoJingWhite}:
\ben
&& E_1 = {{2p/\delta_{sc}(z_1)} \over (1+ (q\nu)^p)} ; \quad
{E_2 \over E_1} = {1+ 2p \over \delta_{sc}(z_1)} + 2e_1; \quad
{E_3 \over E_1} = {4(p^2-1)+6pq\nu \over \delta^2_{sc}(z_1)}+3e_1^2.
\een
For the PS mass function $p=0$, thus all $E_k$ vanishes (also see \citep{Matsubara} for related discussion on non-locally 
biased tracers in both real and redshift spaces in the context of ``Integrated Perturbation Theory'').
%
%
\begin{figure}
\vspace{1.25cm}
\begin{center}
{\epsfxsize=13. cm \epsfysize=4.7 cm 
{\epsfbox[32 425 547 585]{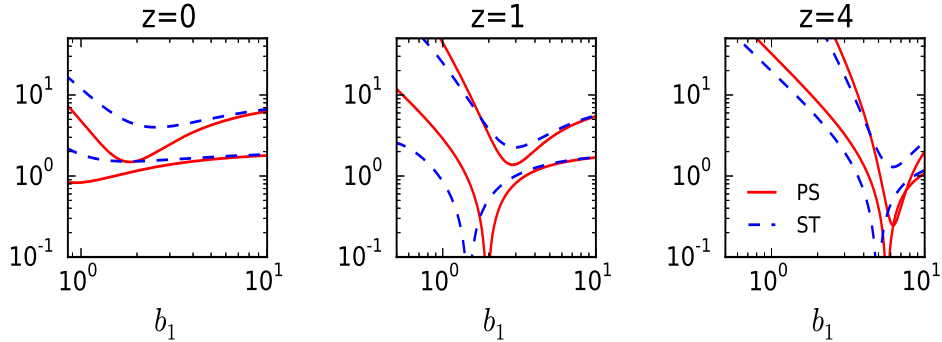}}}
\caption{We plot the normalised CCs ${\mathbb C}^{(h)}_{21}$ and ${\mathbb C}^{(h)}_{31}$ defined respectively
in Eq.(\ref{eq:J21}) and Eq.(\ref{eq:J31}) halos as function of the bias of the halos $b_1$ (defined in Eq.(\ref{eq:b1})).
From left to right the panels correspond respectively to redshifts $z=0$, $z=1$ and $z=4$.
The solid (dashed) lines correspond to the PS (ST) formalism. The halo cumulant correlators
${\mathbb C}^{(h)}_{pq}$ depend on underlying density cumulants ${\mathbb C}_{pq}$ defined in Eq.(\ref{eq:c21})-Eq.(\ref{eq:c41}). We have assumed a locally
power-law power spectrum with a slope $n_{\rm eff}=-1.5$ to evaluate the parameter $\gamma_1$.}
\label{fig:CC_PS_ST}
\end{center}
\end{figure}
%
The cumulants of the halo distribution denoted as ${\mathbb S}^{(h)}_n$can now directly be related to the underlying cumulants ${\mathbb S}_n$ \citep{FryGaztanaga,halo}:
\ben
\label{eq:H3}
&& {\mathbb S}^{(h)}_3 = {1 \over b_1}({\mathbb S}_3 + 3r_2);\\
&& {\mathbb S}^{(h)}_4 = {1 \over b_1^2}({\mathbb S}_4 + 12r_2{\mathbb S}_3 +4r_3 + 12r_2^2); \label{eq:H4}\\
&& {\mathbb S}^{(h)}_5 = {1 \over b_1^3}({\mathbb S}_5 + 20r_2{\mathbb S}_4 + 15r_2{\mathbb S}_3^2 + (30r_3 + 120 r_2^2) {\mathbb S}_3
+ 5r_4 + 60r_3r_2 +60 r_2^3).
\label{eq:H5}
\een
We have introduced the notation $r_k(m) = b_k(m)/b_1(m)$. 
Normalised cumulants for the halos ${\mathbb S}^{(h)}_n$ at a given order depend on all lower order
cumulants of the underlying density field ${\mathbb S}_n$. 
The halo cumulants ${\mathbb S}^{(h)}_3$ depends on $z_1$ and $m$ through the parameters $r_k$.

For small halos, $\nu\ll 1$ identified at early times ($z_1\gg 1$),
$b_1 \approx 1$ and $b_k \approx 0$ for all $k>1$, thus ${\mathbb S}^{(h)}_n={\mathbb S}_n$.
For massive halos $\nu\gg1$ identified at low redshift $b_k=b_1^k$
for $k>1$. Thus in this limit, the halos are completely determined by
the statistical properties of the initial density field. Subsequent
dynamics of gravitational clustering has no effect on their clustering.
For an initial Gaussian random field ${\mathbb S}^{(h)}_n = n^{n-2}$. Thus these
objects follow a lognormal distribution \citep{coles}.

The primary motivation in this section, however, is to derive the CCs 
for the halos and relate them to the IB that will be discussed in the next section.
The lower order halo CCs take the following form:
\ben
\label{eq:J21}
&& {\mathbb C}^{(h)}_{21} = {1 \over b_1}({\mathbb C}_{21}+ 2r_2); \\
\label{eq:J31}
&& {\mathbb C}^{(h)}_{31} = {1 \over b_1^2}({\mathbb C}_{31}+ {7\over 2}r_2 {\mathbb C}_{21} + 6 r_2^2 + 3r_3).
\een
The expressions for underlying ${\mathbb C}_{pq}$ are given in Eq.(\ref{eq:c21})-Eq.(\ref{eq:c41}).
Notice that joint estimation of ${\mathbb C}_{21}$ and ${\mathbb S}_3$ can be used to disentangle $b_1$ and $b_2$.
Later we will see that the CCs for halos and the IB for biased tracers share a similar mathematical structure.

For the peaks the bias parameters $b_k$ follow a similar structure \citep{MoJingWhite}:
\ben
\label{eq:peak_b1}
&& b_1(\nu)=1+{\nu^2+g_1 \over \delta};\\
&& b_2(\nu)= 2(1+a_2){\nu^2 +g_1 \over \delta} + \left ({\nu\over \delta} \right)^2
\left(\nu^2 -1 + 2g_1 +{2g_2 \over \nu_1^2} \right ).
\label{eq:peak_b2}
\een
For the definition of the parameters $g_k$ see Ref.\citep{MoJingWhite}.
The relevant CCs can be derived using the expressions 
Eq.(\ref{eq:J21})-Eq.(\ref{eq:J31}). For $g_k=-1$ the 
results are identical to PS formalism. 

Computation of mass functions in MG theories [ ${\rm F(R)}$ theories ] have been attempted using PS type approach
and involve collapse of spherically symmetric perturbations and can reproduce 
results from numerical simulations \citep{Kopp}. Extension of such results for 
bias parameters computed in parametric MG theories will be presented elsewhere.

For early work on estimation of CCs from the APM galaxy survey see Ref.\citep{SzSz}.
Possible extensions of CCs beyond the two-point function is discussed in Ref.\citep{gen}.

In Figure-\ref{fig:b1b2b3} the first three bias parameters $b_1,\,b_2$ and $b_3$, defined in Eq.(\ref{eq:b1})-Eq.(\ref{eq:b3}), are shown
as a function of halo mass $m$. We have used the ST (thick lines) and PS (thin lines) mass functions defined in Eq.(\ref{eq:ST}).
The plots from left to right correspond respectively to redshifts $z=0, 1$ and $4$.
In Figure-\ref{fig:S3S4} we plot the normalised cumulants ${\mathbb S}^{(h)}_3$ and ${\mathbb S}^{(h)}_4$ defined 
in Eq.(\ref{eq:H3}) and Eq.(\ref{eq:H4}) for halos of different mass as a function of the halo bias $b_1$.
From left to right, the panels correspond respectively to the redshifts $z=0,1$ and $z=4$. The halo cumulants ${\mathbb S}^{(h)}_n$
depend on the underlying density cumulants ${\mathbb S}_n$ Eq.(\ref{eq:S_N}). We have assumed a locally
power-law power law with $n_{\rm eff}=-1.5$ to evaluate the $\gamma_1$ parameter.
We plot the normalised CCs ${\mathbb C}^{(h)}_{21}$ and ${\mathbb C}^{(h)}_{31}$ defined 
in Eq.(\ref{eq:J21}) and Eq.(\ref{eq:J31}) for halos as a function of the bias of the halos $b_1$ in Figure-\ref{fig:CC_PS_ST}.
In addition to the bias parameters, the halo cumulant correlators
${\mathbb C}^{(h)}_{pq}$ depend on underlying density cumulants ${\mathbb C}_{pq}$ Eq.(\ref{eq:c21})-Eq.(\ref{eq:c41})
We have used $n_{\rm eff}=-1.5$ to evaluate the $\gamma_1$ parameter. 
%
%
\begin{figure}
\vspace{1.25cm}
\begin{center}
{\epsfxsize=13. cm \epsfysize=4.7 cm 
{\epsfbox[32 405 547 585]{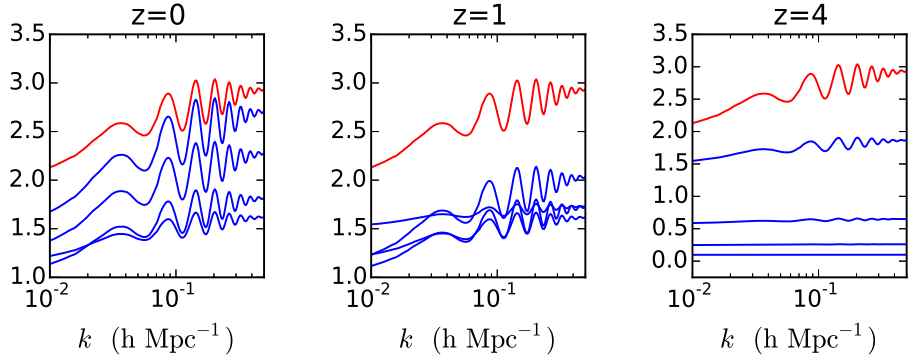}}}
\caption{IB of collapsed objects are plotted as a function the
wave number $k$ for objects with mass $10^{11}{\rm M}_{\odot},10^{12}{\rm M}_{\odot},
10^{13}{\rm M}_{\odot}$ and $10^{14}{\rm M}_{\odot}$. The panels from left to right
correspond to $z=0$, $z=1$ and $z=4$ respectively. We use the expression 
in Eq.(\ref{eq:int_bi}). The bias parameters $b_k(m)$ were computed using the ST mass function
Eq.(\ref{eq:b1})-Eq.(\ref{eq:b3}).
We use halo model based approach to compute the power spectrum.}
\label{fig:IntBI}
\end{center}
\end{figure}
%

\section{Integrated Bispectrum for Biased Tracers in Real Space}
\label{sec:IntBiTri}
The coupling of large and small scale density fluctuations is captured at the lowest order
by the bispectrum in its squeezed limit. The bispctrum is described as a triangular
configuration of three wave vectors. In the squeezed limit on of the wave number
is much smaller than the other two $\bk_1 \ll \bk_2 \approx \bk_3$. In this limit
the bispectrum is effectively collapsed to a power spectrum and can be estimated
by cross-correlating {\em local} density contrast $\delta$ and estimates of {\em local} power spectrum $P_{\delta}(k)$
from survey subvolume. Such cross-correlation effectively bypasses many complications involved in
estimation of the bispectrum (see Ref.\citep{MunshiColes} for relevant discussion and a more complete list of references).
We will derive the squeezed limit of the halo model predicted bispectrum. Later we will extend the discussion to
redshift-space in \textsection{\ref{sec:red}}.

The power spectrum $P_h(k)$ and bispectrum $ {B}_h(\bk_1,\bk_2,\bk_3)$ for the biased tracers are defined as:
\ben
&& \la \delta_h(\bk_1)\delta_h(\bk_2)\ra_c=(2\pi)^3\delta_{\rm 3D}\,(\bk_1+\bk_2)\,P_h(k_1); \\
&& \la \delta_h(\bk_1)\delta_h(\bk_2)\delta_h(\bk_3)\ra_c = (2\pi)^3 \delta_{\rm 3D}(\bk_1+\bk_2+\bk_3)B_h(\bk_1,\bk_2,\bk_3). 
\een                                                                  
Here $\delta_{3\rm D}$ represents 3D Dirac delta function and enforces translational invariance in real space 
(equivalently momentum conservation in Fourier domain). 
Using the bias parameters predicted by the halo model Eq.(\ref{eq:halo_bias}), the power spectrum and bispectrum of halos of a given mass can be 
expressed at tree level in terms of the underlying power and bispectrum \citep{halo}:
\ben
&& {P}_h(\bk) = b^2_1(m)\,{P}_{\delta}(k)\\
&& {B}_h(\bk_1,\bk_2,\bk_3) = b^3_1(m)\,{B}_{\delta}(\bk_1,\bk_2,\bk_3) + 
b_2(m)b_1^2(m) [{P_{\delta}}(k_1){P_{\delta}}(k_2)+ {\rm cyc.perm.}]
\een
We have suppressed the redshift $z$ dependence. The expressions for 
$b_k$ are defined in Eq.(\ref{eq:b1})-Eq.(\ref{eq:b3}) for the halo model. 
The peak theory based bias functions are given in
Eq.(\ref{eq:peak_b1})-Eq.(\ref{eq:peak_b2}). The above expression can be 
generalised to bispectrum of three different types of collapsed objects
having different mass in halo model or peak height in peak theory.
It can be shown that the integrated bispectrum for collapsed objects has the following
expression:
\ben
&&\lim_{\bq_1,\bq_3\rightarrow 0} B_h(\bk-\bq_1, -\bk+\bq_1+\bq_3,-\bq_3)   \nn \\
&&  \stackrel{\text{sq}}{=}
{1\over b_1(m)}\left [ {68 \over 21} -{d\,\ln k^3 {P}_{\delta}(k) \over d\ln k} + 2\,r_2(m) \right ]P_h(k_1)P_h(k_2).
\label{eq:int_bi}
\een
For more discussion on parametrization and derivation of the squeezed limit in Eq.(\ref{eq:int_bi}) see Ref.\citep{MunshiColes}. 
Notice the formal similarity of the  expression in Eq.(\ref{eq:int_bi}) 
and that in Eq.(\ref{eq:c21}).
The expression in Eq.(\ref{eq:int_bi}) consists of two different parts.
The first contribution comes from the underlying mass distribution
with an additional multiplicative factor that depends on linear order bias
$b_1(m)$. The second contribution has a hierarchical nature which depends
on the second order bias coefficient $b_2(m)$.
For $b_1=1$ and $r_1=0$ the expression reduces to that for underlying mass distribution, as expected.
%
%
\begin{figure}
\vspace{1.25cm}
\begin{center}
{\epsfxsize=13. cm \epsfysize=4.7 cm 
{\epsfbox[32 405 547 585]{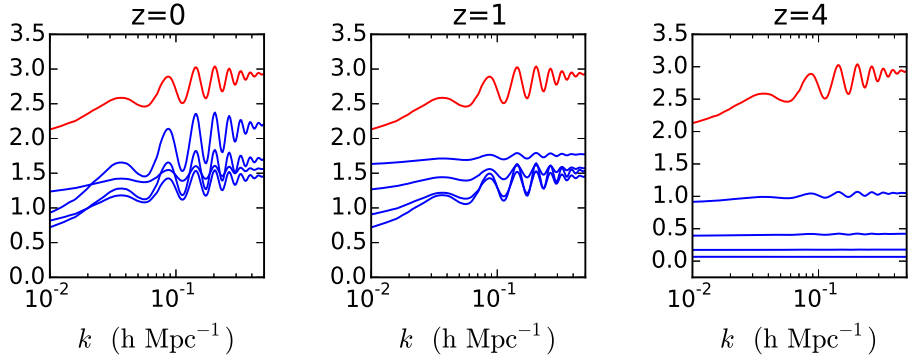}}}
\caption{Same as Figure-\ref{fig:IntBI} but for the PS mass function.}
\label{fig:BI_PS}
\end{center}
\end{figure}

In Figure-\ref{fig:IntBI} and Figure-\ref{fig:BI_PS} we display the result of numerical computation
of the IB as a function of wavenumber $k$. The results in Figure-\ref{fig:IntBI} correspond to the ST model and the ones in Figure-\ref{fig:BI_PS}
correspond to the PS model. The panels from left to right correspond to respectively $z=0,1$ and $z=4$.
Collapsed objects with mass $10^{11}{\rm M}_{\odot},10^{12}{\rm M}_{\odot},
10^{13}{\rm M}_{\odot}$ and $10^{14}{\rm M}_{\odot}$ are considered. The expression 
in Eq.(\ref{eq:int_bi}) is used to compute the bias. The bias parameters $b_k(m)$ were computed using the ST(PS) mass function
and Eq.(\ref{eq:b1})-Eq.(\ref{eq:b3}).

\section{Beyond the halo model: Non-local Eulerian bias}
\label{sec:nonlocal}
The need for more generic biasing model beyond what is predicted by halo model was recently investigated in \citep{McDonaldRoy}.
It was argued that effect of redshift-space distortions aimed at constraining dark energy
or MG theories or neutrino mass will require a more rigorous approach beyond the
halo model as the signatures they leave on the power spectrum are less distinct than the more pronounced Baryonic Acoustic
Osscilations (BAO) signatures.
Modification of local deterministic bias predicted by the halo model can be made to
include terms that are stochastic, non-local terms. In this section, we extend the 
generating function based approach introduced in \textsection\ref{sec:cc} to more general biasing schemes.

``Non-local Eulerian bias'' was proposed in Ref.\cite{McDonaldRoy} and was recently investigated in Ref.\cite{GilMarin}, semi-analytical modelling
or a combination of both. It generalises the deterministic local halo bias presented in Eq.(\ref{eq:halo_bias}):
\ben
&& \delta_h = b_{\delta} \delta +{1 \over 2!}b_{\delta^2}(\delta^2 - \la \delta^2\ra)
+ {1 \over 2!}b_{s^2} \left(s^2 -\la s^2\ra \right ) \nn \\
&& \quad\quad + {1 \over 3!} b_{\delta^3} \delta^3 + {1 \over 2!} b_{\delta s^2}\delta s^2 
 + b_{\psi} \psi + b_{st} st + {1\over 3!} b_{s^3}s^3 + \cdots.
\label{eq:non_local}
\een
The notations introduced in Eq.(\ref{eq:non_local}) are defined below:
\ben
\label{eq:tidal}
&& s_{ij}(\bx)\equiv\partial_i\partial_j \Phi(\bx) -{1\over 3}\delta^{\rm K}_{ij} \delta(\bx); \quad  \triangle \Phi = \delta;\\
&& s_{ij} \equiv \gamma_{ij}\delta({\bf x});\quad  t_{ij}({\bf x}) \equiv  \gamma_{ij}\eta(\bx); \quad \eta= \theta(\bx)-\delta(\bx);\\
&& \psi(\bx) \equiv \eta(\bx) -{2\over 7}s^2(\bx) +{4 \over 21}\delta^2(\bx); \quad 
\gamma_{ij} = \left [ \partial_i\partial_j \nabla^{-2} -{1 \over 3}\delta_{ij} \right ].
\label{eq:gamma}
\een
Here, $\Phi$ represents the gravitational field,  $s$ the tidal tensor, $\theta$ the divergence of velocity and $\delta^{\rm K}_{ij}$ denotes
the Kronecker delta function. 
The difference variables $\eta$ and $t$ are only at second order. The variable $\psi$ is non-zero only at third order
and thus do not contribute at the level of bispectrum. By construction, the tensors $s$ and $t$ are traceless.  
We have introduced the following short-hand notations above :
\ben
s^2(\bx) \equiv s_{ij}(\bx)s_{ij}(\bx); \quad\quad
s^3(\bx) \equiv s_{ij}(\bx)s_{jk}(\bx)s_{ki}(\bx); \quad\quad st \equiv s_{ij}(\bx)t_{ji}(\bx).
\label{eq:s2}
\een
Summation over the repeated indices is assumed.

The parameters $b_{\delta}, b_{\delta^2}, b_{s^2}, \cdots$ will be left arbitrary. In this section we will show that
such generic biasing scheme can also be handled using the generating function approach introduced in \textsection{\ref{sec:cc}}.
We will also compute the IB for such a biasing scheme.

Gravitational collapse introduces non-locality and non-linearity.
The above model incorporates both non-linear term involving $\delta^2(\bx)$ and the 
non-local $s^2(\bx)\equiv s_{ij}(\bx)s^{ij}(\bx)$ terms ($s({\bf x})$ represents the tidal tensor defined in Eq.(\ref{eq:tidal}) below)\citep{McDonaldRoy}. 
The presence of terms $\langle\delta^2\rangle$ 
and $\langle s^2\rangle$ ensures $\la\delta_h\ra=0$. These term contribute to power spectrum at one-loop level but make tree-level
contributions to the bispectrum and hence need to be included even when analysing large scale data
using perturbative results. 

Using the properties of the generating function listed in Eq.(\ref{eq:gen_1}) we can relate the generating function
${\cal G}_{\delta_h}$ of galaxies to generating functions ${\cal G}_{\delta}, {\cal G}_{s}, \cdots$:
\ben
&& {\cal G}_{\delta_h} = b_{\delta}  {\cal G}_\delta +
{1 \over 2!}b_{\delta^2}{\cal G}_\delta^2 
+ {1 \over 2!} b_{s^2}{\cal G}_{s^2} \nn \\
&& \quad\quad +{1 \over 3!} b_{\delta^3}{{\cal G}}^3_\delta + {1 \over 2!} b_{\delta s^2}{\cal G}_{\delta}
{{\cal G}}_{s^2}
 + b_{\psi}{\cal G}_\psi + {1\over 3}b_{st}{\cal G}_{st} + {1\over 3!} b_{s^3}{\cal G}_{s^3} + \cdots.
\label{eq:G_halo}
\een
Taylor expanding  ${\cal G}_{\delta_h}$ can provide us the relations that will generalise Eq.(\ref{eq:cc1})-Eq.(\ref{eq:cc4}) to {\em arbitrary} order.

Using Eq.(\ref{eq:tidal})-Eq.(\ref{eq:gamma}) various generating functions that appear in Eq.(\ref{eq:G_halo})
can be expressed in terms of the generating functions ${\cal G}_{\delta}$ and ${\cal G}_{\theta}$.
\ben
&& {\cal G}_{s^2}(\tau)= 0; \quad {\cal G}_{\triangle \Phi}(\tau) = {\cal G}_{\delta}(\tau); \quad {\cal G}_{\eta}(\tau) = {\cal G}_{\delta}(\tau)-{\cal G}_{\theta}(\tau); 
\quad {\cal G}_{\psi}(\tau) = {\cal G}_{\eta}-{4\over 21}{\cal G}^2_{\delta}.
\een
Thus, ${\cal G}_{h}$ depends only on the bias coefficients and the generating functions ${\cal G}_{\delta}, {\cal G}_{\theta}$.
This implies that the halo moments, at all order, in such biasing models, can be expressed in terms of the $\delta$ and $\theta$ moments.
Indeed, ${\cal G}_{s^2}(\tau)= 0$ implies the Eq.(\ref{eq:J21}) derived for halo models, using a local biasing scheme in Eulerian space, remains
unchanged. Similar results will hold for the skewness paramter Eq.(\ref{eq:H3}). However the power spectrum
will change due to loop-level corrections \citep{Beutler}.

Assuming galaxy bias to be local in Lagrangian space it is possible to express the non-local bias term 
in terms of the linear bias e.g. $b_{s^2}=-{4/7}(b_{\delta}-1)$ \citep{Chan,Baldauf}. Similar results exist relating e.g $b_{\delta^3}$ and $b_{\delta}$ \citep{Saito}.
The bispectrum depends only on the parameters $b_{\delta},b_{\delta^2}$ and $b_{s^2}$ at the {\em tree-level}. In recent studies
the second order bias parameter $b_{\delta^2}$ was found to be sensitive to the truncation effect due to the presence of
higher order terms. In many studies, it is also considered to be a nuisance parameter \citep{GilMarin_bias}. 

With this replacement in Eq.(\ref{eq:cc1})-Eq.(\ref{eq:cc4}), the resulting
$\nu$ s can be used in Eq.(\ref{eq:S_N})-Eq.(\ref{eq:S_Np}) to compute the ${\mathbb S}_n$ parameters
in this model. For CCs the relevant equations are Eq.(\ref{eq:J21})-Eq.(\ref{eq:J31}).

Next, to compute the IB we use the following model of the bispectrum \citep{GilMarin}:
\ben
&& P_h(k) = b_\delta^2 P(k); \\
&& B_h(\bk_1,\bk_2,\bk_3)= b_\delta^3B_\delta(\bk_1,\bk_2,\bk_3) + \nn \\
&& \quad\quad\quad b_\delta^2 \left [b_{\delta^2} P_{\delta}(\bk_1)P_{\delta}(\bk_2) +b_{s^2}P_{\delta}(\bk_1)P_{\delta}(\bk_2)S_2(\bk_1,\bk_2) 
+ {\rm cyc.perm.} \right ]; \\
&& S_2(\bk_1,\bk_2) \equiv {(\bk_1\cdot \bk_2)^2 \over (k_1k_2)^2} - {1 \over 3}.
\een
Here $P_{\delta}(k_1)$ and $B_\delta(\bk_1,\bk_2,\bk_3)$ are matter power spectrum and bispectrum respectively while $S_2(\bk_1,\bk_2)$ represents the kernel that is used to express the Fourier transform of $s^2({\bf x})$ defined in Eq.(\ref{eq:s2}):
\ben
&& s^2({\bk}) =  \int{d^3\bk^{\prime} \over (2\pi)^3} S_2(\bk^{\prime},\bk-\bk^{\prime})\delta(\bk^{\prime})\delta(\bk-\bk^{\prime}).
\een
It can be shown that in the squeezed limit  $S_2(\bk_1,\bk_2)$ 
vanishes and the final results are independent of $b_{s^2}$:
\ben
&&\lim_{\bq_1,\bq_3\rightarrow 0} B_h(\bk-\bq_1, -\bk+\bq_1+\bq_3,-\bq_3)   \nn \\
&&  \stackrel{\text{sq}}{=} \quad\quad {1\over b_{\delta}}\left [\left ({68 \over 21}- {1 \over 3}(n_{\rm eff}+3) \right ) + 2{b_{\delta^2}\over b_{\delta}}\right ]
P_h(k_1)P_h(q_3).
\een
We have assumed a power spectrum that can be approximated as a power law $P_{\delta}(k)\propto k^{n_{\rm eff}}$.
The bispectrum is assumed to be a tree-level bispectrum. Its performance beyond the perturbative regime
can substantially be improved by substituting the kernel $F_2(\bk_1,\bk_2)$ with an effective kernel which depend on
parameters that are calibrated using N-body simulation; e.g. Ref.\citep{GilMarin_bias} introduced a set of nine parameters
each for the kernels $F_2(\bk_1,\bk_2)$ and $G_2(\bk_1,\bk_2)$ that defines the bispectrum of $\delta(\bx,a)$ and $\theta(\bx,a)$
respectively (to be defined later in \textsection\ref{sec:red} in Eq.(\ref{eq:F2})).
%
%
\section{The Bispectrum and Redshift-Space Distortion}
\label{sec:red}
%
%
Many authors have contributed to the development of the theory galaxy bispectrum \citep{MVH1, Scocci1, Verde1, Scocci2, Scocci3, HMV2}
and its estimation from survey data \citep{FFFS,SFFF,JingBorner,GazScoc,WangYang,Marin}.

Using the Eulerian perturbation theory (EPT), the density contrast $\delta(\bx,\tau)$
and the divergence of velocity $\theta(\bx,\tau)$ are expressed as a sum of perturbative terms
$\delta(\bx,\tau)=\sum\delta^{(n)}(\bx,\tau)$ and $\theta(\bx,\tau)=\sum\theta^{(n)}(\bx,\tau)$.
In our notation ${\bf v} = d{\bf x}/d\tau$ is the peculiar velocity in comoving coordiante ${\bf x}$ w.r.t. conformal time $\tau$ 
and  $\theta \equiv \nabla\cdot {\bf v}$ is the comoving velocity divergence.
In the Fourier domain, the $n^{\rm th}$ order perturbative expressions 
for the density contrast $\delta^{(n)}(\bk,\tau)$  can be expressed in
terms of convolution of kernel,
$F_n(\bk_1,\cdots,\bk_n)$ or $G_n(\bk_1,\cdots,\bk_n)$, and the linear density contrast $\delta^{(1)}(\bk_i)$
(see Ref.\citep{review} for detailed derivations and related discussions):
\ben
\delta(\bk,\tau) = \sum^{\infty}_{n=1} D^n_+(\tau)\,[\delta_{\rm D}]_{n}\,\int d^3\bk_1\cdots \int d^3\bk_n\, F_n(\bk_1,\cdots,\bk_n)
\,\delta^{(1)}(\bk_1)\cdots\delta^{(1)}(\bk_n).
\een
Similar expression holds for the Fourier transform of the divergence of velocity $\theta(\bk,\tau)$. The corresponding kernel will be denoted as 
$G_n(\bk_1,\cdots,\bk_n)$. In our notation, $D_+(\tau)$ is the linear growth factor.

For the computation of integrated bispectra we will only require the following
{\em symmetrized} second order kernels $F_2(\bk_1,\bk_2)$ and $G_2(\bk_1,\bk_2)$ that have the following forms \citep{MunshiColes}:
\ben
&& F_2(\bk_1,\bk_2) = {1\over 2}{(1+\epsilon)} + {1 \over 2}{\mu_{12}}\left ( {k_1\over k_2}+{k_2\over k_1}\right )
+{1\over 2}{(1-\epsilon)}\mu^2_{12}; \quad \mu_{12}={\hat \bk_1\cdot \hat \bk_2}.
\label{eq:F2}
\een
The symmetrized kernels are constructed by taking the mean of all possible permutations of the
unsymmetrized kernels. Here $[\delta_{\rm D}]_n=\delta_{\rm 3D}(\bk-\bk_1\cdots-\bk_n)$ represents the momentum conserving Dirac delta function in 
three dimensions (3D). The parameter takes the value $\epsilon=3/7$ for an Einstein de-Sitter Universe. The kernel $G_2(\bk_1,\bk_2)$ 
has similar functional form. We will use a different parameter $\epsilon^{\prime}$ to avoid 
confusion for $\theta$ with $\epsilon^{\prime}=-1/7$. 
For the first order Lagrangian Perturbation Theory (LPT), also known as the Zel'dovich Approximation (ZA; see e.g. 
Ref.\citep{MunshiStarobinsky,MunshiSahniStarobinsky}),
the parameter takes the value $\epsilon={0}$.

Redshift-space distortions are result of peculiar velocities of galaxies which arise due
to gravitational clustering. These distortions are a known source of complication
in interpretation of clustering from spectroscopic surveys. 
The redshift-space density contrast $\delta_{h,s}$ of halos
can be expressed in terms of redshift-space kernels $Z_n(\bk_1,\cdots,\bk_n)$ and the bias parameters $b_k$:
\ben
\delta_{h,s}(\bk,\tau) = \sum^{\infty}_{n=1} D^n_+(\tau) [\delta_D]_n \int d^3\bk_1\cdots \int d^3\bk_n\, Z_n(\bk_1,\cdots,\bk_n)
\delta_L(\bk_1)\cdots\delta_L(\bk_n).
\een
The lower order kernels $Z_1(\bk_1)$ and $Z_2(\bk_1,\bk_2)$ in redshift-space are given by the following expressions \citep{red_roman}:
\ben
\label{eq:Z1}
&& Z_1(\bk_i) \equiv (b_1+f\mu_i^2) \\
&& Z_2(\bk_1,\bk_2)\equiv b_1\left [F_2(\bk_1,\bk_2)+ 
{1 \over 2}f\mu k\left ({\mu_1 \over k_1} + {\mu_2 \over k_2} \right )\right ]
+ f\mu^2 G_2(\bk_1,\bk_2\nn) \label{eq:Z2}\\
&& \quad\quad\quad + {1 \over 2}f\mu k \mu_1\mu_2\left ({\mu_1 \over k_1} + {\mu_2 \over k_2} \right )
+{b_2 \over 2} +{b_{s^2} \over 2}S_2(\bk_1,\bk_2).
\een
Here, $f = d \ln D_+/d \ln a$ is the logarithmic growth rate and $a$ is the scale factor. 
The kernel  $Z_2(\bk_1,\bk_2)$ defined in Eq.(\ref{eq:Z2}) depends on both $F_2(\bk_1,\bk_2)$ and $G_2(\bk_1,\bk_2)$ which implies that the squeezed limit of redshift
space bispectrum will depend on the squeezed limits of $\delta$ and $\theta$ bisepctrum.
To relate the real-space density contrast of halos $\delta_h$ to the underlying density contrast $\delta$
we use a deterministic bias: $\delta_h= \sum_k b_k {\delta^k / k!}$.

The FoG effect arises as a result of random peculiar velocities of galaxies within virialised collapsed objects.
The effect of peculiar velocity is an incoherent contribution and results in a suppression
of the clustering amplitude at high $k$ \citep{Jack72}. Throughout, we have used the following expressions:
\ben
&& \mu =  {\hat \bx}_{\parallel} \cdot {\hat \bk};\quad\quad \mu_i ={\hat \bx}_{\parallel}\cdot{\hat\bk}_i; \\
&& \bk=\bk_1+\bk_2; \quad \mu k \equiv (\mu_1 k_1 + \mu_2 k_2); \quad \bk^2 = (\bk_1+\bk_2)^2. 
\een
The hats represent unit vectors i.e. ${\hat \bk}={\bf k}/|\bk|$ and  ${\hat \bk_i}={\bf k}_i/|\bk|$. 
The comoving separation separated into components that are parallel and perpendicular to the 
line of sight $\bx=\bx_{\parallel}+\bx_{\perp}$ and $\hat \bx_{\parallel}= \bx_{\parallel}/|\bx_{\parallel}|$ is the unit vector along the line-of-sight.
In the redshift-space the halo power spectrum takes the following form \citep{review}: 
\ben
&&  \langle\delta_{h,s}(\bk_1)\delta_{h,s}(\bk_2)\rangle \equiv P_{h,s}(k_1) \delta_{\rm 3D}(\bk_1+\bk_2); \\
&& P_{h,s}(k) =  b_1^2(1+ b_1^{-1}f\mu_k^2)^2\,P_{\delta}(k).
\een
The halo bispectrum in redshift-space has the following expression \cite{red_roman}:
\ben
&& \langle\delta_{h,s}(\bk_1)\delta_{h,s}(\bk_2)\delta_{h,s}(\bk_3)\rangle_c \equiv
\delta_{\rm 3D}(\bk_1+\bk_2+\bk_3)B_{h,s}(\bk_1,\bk_2,\bk_3);\\
&& B_{h,s}(\bk_1,\bk_2,\bk_3) = D_{\rm FoG}
[2P(\bk_1){\rm Z}_1(\bk_1)P(\bk_2){\rm Z}_1(\bk_2){\rm Z}_2 (\bk_1,\bk_2) + {\rm cyc.perm.}].
\label{eq:bispec_red}
\een
The contribution from the finger-of-god (FoG) appears as a multiplicative factor:
\ben
&& D_{\rm FoG}(k_1,k_2,k_3,\sigma_{\rm FoG}[z])
=(1+[k_1^2 \mu^2_{1} + k_2^2 \mu^2_{2} + k_3^2 \mu^2_{3}]^2 \sigma^2_{\rm FoG}[z]/2)^{−2}.
\een

The various configurations to the bispectrum in Eq.(\ref{eq:bispec_red}) can be grouped into the following categories \citep{Chinag_thesis}:
\ben
&& {B} = {B}_{\rm SQ_1}+{B}_{\rm SQ_2}+{B}_{\rm NLB}+{B}_{\rm FoG}; \\
&& {B_{SQ_1}} = b_1^3\sum \beta^{ i-1}B_{{SQ_1},i}; \quad {B_{SQ_2}} = b_1^3\beta \sum \beta^{ i-1}B_{{\rm SQ_2},i};\\
&& {B_{\rm NLB}} = b_1^2b_2\beta \sum \beta^{ i-1}B_{{\rm SQ_2},i}. \\
&& {B}_{\rm FoG} = b_1^4 \beta[ B_{\rm FoG_1}+\beta (B_{\rm FoG_2} +B_{\rm FoG_3}) +
\beta^2(B_{\rm FoG_4}) + \beta^3 B_{\rm FoG_6}].
\een
Here, $\beta = {f/b_1}$. The contributions ${B}_{\rm SQ_1}$ represents linear squashing and depends on the kernel $F_2(\bk_1,\bk_2)$.
The linear Kaiser effect \citep{Kaiser87} which represents the coherent distortion of the peculiar velocity
along the LoS. Its effect is controlled by the linear growth rate. At the level of power spectrum it leads to
an enhancement of the power spectrum amplitude at small $k$.
The second order squashing terms  ${B}_{\rm SQ_2}$ depends on the kernel $G_2(\bk_1,\bk_2)$.
Nonlinear biasing is represented by the  $B_{\rm NLB}$ terms and finally $B_{\rm FOG}$
represents the FoG effect. The squeezed limits of these contributions
will be presented in \textsection\ref{sec:red_IB}.

To recover the real-space expressions we need to take the limit $f\rightarrow 0$ in Eq.(\ref{eq:Z1})-Eq.(\ref{eq:Z2}).
The above expressions are valid in the flat-sky approximation. Future surveys will probe a 
considerable fraction of the sky. A 3D approach has been developed that 
uses spherical-Bessel transform has been used recently the redshift power spectrum 
\citep{PM13,galaxy_red, Bernardeau_wideangle}.

\section{Bispectrum in Parametrized Modified Gravity Theories : $\gamma$ Models}
\label{sec:MG}
%
To illustrate our approach we will compute the IB for a class phenomenological models of MG theories
developed in Ref.\citep{Bernardeau_Brax} - the so called $\gamma$ models.
These particular set of models were constructed by modifying the Euler equation
of the Continuity-Euler-Poisson system.
For these parametrization described  in Ref.\citep{Bernardeau_Brax} the $\epsilon$ and $\epsilon^{\prime}$
defining the kernels $F_2(\bk_1,\bk_2)$ and $G_2(\bk_1,\bk_2)$ can be expressed in terms of time dependent parameters $\nu_2$ and $\mu_2$:
\ben
\epsilon={3 \over 2}\nu_2-2;  \quad \epsilon^{\prime}={3 \over 2}\mu_2-2.
\label{eq:numu}
\een
The parameters  $\hat\nu_2$ and $\hat\mu_2$ are defined through the following fitting formulae \citep{Bernardeau_Brax}:
\ben
&& \nu_2(\gamma) = \nu_2^{\rm GR} - {10 \over 273}(\gamma -\gamma^{\rm GR})(1-\Omega_{\rm M})\Omega_{\rm M}^{\gamma^{\rm GR}-1}; \\
&& \mu_2(\gamma) = \mu_2^{\rm GR} - {50 \over 273}(\gamma -\gamma^{\rm GR})(1-\Omega_{\rm M})\Omega_{\rm M}^{\gamma^{\rm GR}-1}.
\een
where the parameters $\nu_2^{\rm GR}$ and  $\mu_2^{\rm GR}$ are defined by the following relations \citep{Bernardeau_Brax}:
\ben
&& \nu_2^{\rm GR} = {4 \over 3} +{2 \over 7}{\Omega_{\rm M}^{-1/143}}; 
\quad \mu_2^{\rm GR} = -{4 \over 21} +{10 \over 7}{\Omega_{\rm M}^{-1/143}};
\een
\n
For GR and $\Omega=1$ we have, $\nu^{\rm GR}_2={34/21}$ and $\mu^{\rm GR}_2 = {26/21}$. Using Eq.(\ref{eq:numu}) we recover  $\epsilon={3/7}$ and $\epsilon^{\prime}=-{1/7}$.
The growth factor in these models scales as: $f = {d\ln D_+/d\ln a} \approx \Omega_M^{\gamma}$ with $\gamma_{\rm GR}=6/11\approx 0.55$ for GR and for DGP braneworld  models $\gamma=0.67$. 
The method developed below can be extended to other parametric MG theories. 

Notice that the {\em unsmoothed} skewness parameter ${\mathbb S}_3= 3\nu_2$ and ${\mathbb T}_3\equiv {\la\theta^3 \ra/_c\la\theta^2\ra_c^2}= 3\mu_2$. 
The lowest order CC, ${\mathbb C}_{21}$,  for $\delta$ is given by ${\mathbb C}_{21}=2\nu_2$.

\section{Integrated Bispectrum in Modified Gravity Theories}
\label{sec:red_IB}
In a squeezed configuration, the bispectrum effectively describes the effect of long wavelength modes 
on small scale structures. This corresponds to the lowest order coupling of small and long wavelength modes
and has been studied in the literature by many authors recently ( see \citep{Chinag_thesis} and \citep{MunshiColes} the references therein).

In the squeezed limit the wave vectors can be parametrize as: $\bk_1=\bk-\bq_a$; $\bk_2=-\bk+\bq_1+\bq_2$ and $\bk_3=-\bq_3$.
Taylor expanding the wave-vectors and keeping only the linear order terms in $q_a$ and $q_b$ we arrive at:
\ben
&& k_1 = k\left ( 1- \mu_{ak}{q_a \over k}\right ); \quad
k_2 = k\left ( 1- \mu_{ak}{q_a \over k}-\mu_{bk}{q_b \over k} \right ); \quad 
k_3 = -q_3.
\een
Taylor expanding the power spectrum and retaining only the linear order in $q_a$ and $q_b$:
\ben
&& P_{\delta}(k_1) = P_{\delta}(k)\left [1 -{q_a\mu_{ak} \over k}{d\ln P_{\delta}(k) \over d\ln k}\right ]; \;\;\\
&& P_{\delta}(k_2)=P_{\delta}(k)\left [1 -{1 \over k}({q_a\mu_{ak} +q_b\mu_{bk}}){d\ln P_{\delta}(k) \over d\ln k}\right ];\;\; P_{\delta}(k_3)=P_{\delta}(q_b).
\een
Following notations are used to represent the cosines of various angles:
\ben
\mu_k = \hat {\bf k}\cdot\hat {\bf x}_{\parallel}; \quad \mu_a = \hat {\bf q}_a\cdot\hat {\bf x}_{\parallel};
\quad \mu_b = \hat{\bf q}_b\cdot \hat {\bf x}_{\parallel}; \quad \mu_{ab} =\hat {\bf q}_a\cdot \hat {\bf q}_b.
\label{eq:gen}
\een
Where $\hat{\bf k} = {\bf k}/|k|$. The angular variables $\mu_i$ can similarly be expressed as:
\ben
&& \mu_1 = \mu_k + {1 \over k}\left (q_a \mu_{ak}\mu_k -q_a\mu_a\right )+\cdots;\;\;\\
&& \mu_2 = -\mu_k + {1 \over k}\left (q_a\mu_k+q_b\mu_{bk} -q_a\mu_a\mu_k -q_b\mu_b\mu_k \right )+\cdots; \nn \\
&& \hspace{0.5cm} = \mu_1 + {1\over k}\left ( q_b\mu_{bk} -q_b\mu_b\mu_k \right)+\cdots; \\
&& \mu_3 = -\mu_b.
\een
We have expanded the variables $\mu_1,\mu_2$ and $\mu_3$ in Taylor series and kept terms up to linear order in $(q_a/k)$ and $(q_b/k)$. 
In the squeezed limit $\bk_1=\bk-\bq_a$; $\bk_2=-\bk+\bq_1+\bq_2$ and $\bk_3=-\bq_3$, 
and the perturbative kernel $F_2(\bk_1,\bk_2)$, defined in Eq.(\ref{eq:F2}), takes
the following form:
\ben
\label{eq:F1}
&& F_2(\bk_1,\bk_2)=0; \\
&& F_2(\bk_1,\bk_3)= {1 \over 2}(1+\epsilon) - {1 \over 2q_b}({k\mu_{bk}-q_a\mu_{ab}}) + {1 \over 2}(1-\epsilon)\mu^2_{bk} ; \\
&& F_2(\bk_2,\bk_3)= {\epsilon\over 2}+{1 \over 2q_b}({k\mu_{bk}-q_a\mu_{ab}}) + {1\over 2}(1-\epsilon)\mu^2_{bk}.
\label{eq:F3}
\een
For GR we have $\epsilon=3/7$, while for the Zel'dovich Approximation (ZA) we have $\epsilon={1/2}$.
Similar results hold for $G_2$
where the parameter with $\epsilon^{\prime}$ for GR has the numerical value $\epsilon^{\prime}=-1/7$.
Indeed for the $\gamma$-models of \citep{Bernardeau_Brax} $\epsilon$ is independent of $k$ but for more generic class of models 
(see e.g. Ref.\citep{BraxValageas12}) this will not be the case; a dedicated analysis will be presented elsewhere.
 
\begin{figure}
\vspace{1.25cm}
\begin{center}
{\epsfxsize=13. cm \epsfysize=4.7 cm 
{\epsfbox[32 405 547 585]{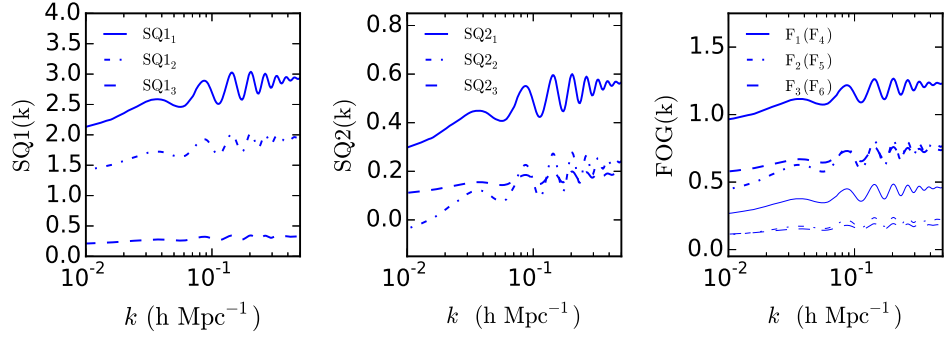}}}
\caption{Various contributions to the IB are displayed as a function of the wavenumber $k$. The left panel shows
the contributions from the terms Eq.(\ref{eq:sq1A})-Eq.(\ref{eq:sq1C}) middle panel correspond 
to Eq(\ref{eq:sq2A})-Eq.(\ref{eq:sq2C}). The right panel corresponds to contribution from Eq.(\ref{eq:FOGA})-Eq.(\ref{eq:FOGF}).
The power spectrum is approximated locally as a power law with a logarithmic slope of $n$ that reduces the 
expressions to the form Eq.(\ref{eq:gen}) with coefficients given by Eq.(\ref{eq:SQ1,1})-Eq.(\ref{eq:SQ2,3}).
Results are shown for $z=0$.}
\label{fig:redshift}
\end{center}
\end{figure}

We will consider the ``linear squashing'' terms (SQ$_1$) that depend on the kernel $F_2(\bk_1,\bk_2)$ first \citep{Chinag_thesis}:
\ben
&& B_{\rm SQ1_1}(\bk_1,\bk_2,\bk_3) = 2[F_2(\bk_1,\bk_2)P_{\delta}(k_1)P_{\delta}(k_2)+{\rm cyc.perm.}];\\
&& B_{\rm SQ1_2}(\bk_1,\bk_2,\bk_3) = 2[(\mu_1^2+\mu_2^2)F_2(\bk_1,\bk_2)P_{\delta}(k_1)P_{\delta}(k_2)+{\rm cyc.perm.}];\\
&& B_{\rm SQ1_3}(\bk_1,\bk_2,\bk_3) = 2[\mu_1^2\mu_2^2F_2(\bk_1,\bk_2)P_{\delta}(k_1)P_{\delta}(k_2)+{\rm cyc.perm.}].
\een
For a parametrization similar to defined in Eq.(\ref{eq:F2}) we have the following expressions:
\ben
\label{eq:sq1A}
&& B_{\rm SQ1_1} \stackrel{\text{sq}}{=} \left [\left( {1} +2\epsilon\right ) + 2(1-\epsilon)\mu^2_{bk} -{d\ln P_{\delta}(k) \over d\ln k}\mu^2_{bk} \right ]\pfact;\\
&& B_{\rm SQ1_2} \stackrel{\text{sq}}{=} \Big [{\left( {1} + 2\epsilon \right )}\mu_k^2+{2(2-\epsilon)}\mu^2_{bk}\mu_k^2 
-2 \mu_{bk}\mu_k \mu_b + \left({1}+2\epsilon\right)\mu_b^2 +{2(1-\epsilon)}\mu^2_{bk}\mu_b^2\nn \\
&& \hspace{3cm} -{d\ln P_{\delta}(k) \over d\ln k}\left( \mu^2_{bk}\mu_k^2 +\mu^2_{bk}\mu_b^2\right )   \Big ]\pfact; \\
&& B_{\rm SQ1_3} \stackrel{\text{sq}}{=}  \Big [ \left ({1}+2\epsilon \right ) \mu_k^2\mu_b^2+ 2(2-\epsilon) \mu^2_{bk}\mu_k^2\mu_b^2-2 \mu_{k}\mu_k\mu_b^3 \nn \\
&& \hspace{3cm} - {d\ln P_{\delta}(k) \over d\ln k}\mu^2_{bk}\mu^2_{k}\mu^2_b \Big ]\pfact.
\label{eq:sq1C}
\een
The other set of terms are the ``second order squashing'' terms (SQ$_2$) and depend on the kernel $G_2(\bk_1,\bk_2)$ \citep{Chinag_thesis}:
\ben
&& B_{\rm SQ2_1} = 2[\mu^2G_2(\bk_1,\bk_2)P_{\delta}(k_1)P_{\delta}(k_2)+{\rm cyc.perm.}];\\
&& B_{\rm SQ2_2} = 2[\mu^2(\mu_1^2+\mu_2^2)G_2(\bk_1,\bk_2)P_{\delta}(k_1)P_{\delta}(k_2)+{\rm cyc.perm.}];\\
&& B_{\rm SQ2_3} = 2[\mu^2\mu_1^2\mu_2^2G_2(\bk_1,\bk_2)P_{\delta}(k_1)P_{\delta}(k_2)+{\rm cyc.perm.}].
\een
In the squeezed limit they take the following form:
\ben
\label{eq:sq2A}
&& B_{\rm SQ2_1} \stackrel{\text{sq}}{=} \left[\left ({1}+2\epsilon^{\prime} \right )\mu_k^2  -2{\epsilon^{\prime}}\mu^2_{bk}\mu^2_k + 2\mu_{bk}\mu_k\mu_b
-{d\ln P_{\delta}(k) \over d \ln k}\mu^2_{bk}\mu^2_k\mu^2_b\right ]\pfact; \\
&& B_{\rm SQ2_2} \stackrel{\text{sq}}{=} \Big [ {(1+2\epsilon^{\prime})}\mu_k^4 + 2(1-\epsilon^{\prime})\mu_{bk}^2\mu_k^4 +{(1+2\epsilon^{\prime})}\mu_k^2\mu_b^2
-2\epsilon^{\prime}\mu_{bk}^2\mu_k^2\mu_b^2+ 2\mu_{bk}\mu_k\mu_b^3 \nn \\
&& \hspace{3cm}-{d\ln P_{\delta}(k) \over d\ln k}(\mu_{bk}^2\mu_b^4+\mu^2_{bk}\mu_k^2\mu_b^2) \Big ]\pfact;\\
&& B_{\rm SQ2_3} \stackrel{\text{sq}}{=} \left [(1+2\epsilon^{\prime})\mu_k^4\mu_b^2 +2(1-\epsilon^{\prime})\mu^2_{bk}\mu_k^4\mu_b^2 - 
{d\ln P_{\delta}(k)\over d\ln k}\mu^2_{bk}\mu_k^4\mu_b^2 \right ]\pfact.
\label{eq:sq2C}
\een

The following three contributions correspond to the ``non-linear bias'' (NLB) terms and are independent of kernels $F_2(\bk_1,\bk_2)$ and
$G_2(\bk_1,\bk_2)$, so do not depend on $\epsilon$ or $\epsilon^{\prime}$. In the squeezed limit they take the following form \cite{Chinag_thesis}: 
\ben
&& B_{\rm NLB,1} \stackrel{\text{sq}}{=} 2\pfact;\quad\\
&& B_{\rm NLB,2} \stackrel{\text{sq}}{=} 2\left [\mu_k^2 +\mu_b^2 \right ]\pfact;\quad\\
&& B_{\rm NLB,3} \stackrel{\text{sq}}{=} 2\left [\mu_k^2\mu_b^2 \right ]\pfact.
\een
The contribution from the ``Finger-of-God'' (FOG) effect in the squeezed limit \cite{Chinag_thesis} are given by:
\ben
\label{eq:FOGA}
&& B_{\rm FOG_1} \stackrel{\text{sq}}{=} \left [ 2\mu_k^2 +\mu_b^2 -{d\ln P_{\delta}(k) \over d\ln k}\mu_{bk}\mu_k\mu_b \right ]\pfact;\\
&& B_{\rm FOG_2}\stackrel{\text{sq}}{=} \left [ 4\mu_{bk}\mu_k^3\mu_b + 2\mu_k^2\mu^2_b - 
2{d\ln P_{\delta}(k)\over d\ln k}\mu_{bk}\mu_k^3\mu_b \right ]\pfact;\\
&& B_{\rm FOG_3}\stackrel{\text{sq}}{=} \left [2\mu_k^4 + \mu_b^4 - {d\ln P_{\delta}(k)\over d\ln k}\mu_{bk}\mu_k\mu_b^3 \right ]\pfact;\\
&& B_{\rm FOG_4}\stackrel{\text{sq}}{=} \left [ 4\mu_k^4 \mu_b^2 + 4\mu_{bk} \mu_k^3 \mu_b^3 - 2\mu_k^2 \mu_b^4 - 
2{d\ln P_{\delta}(k)\over d\ln k} \mu_{bk} \mu_k \mu_b \right ]\pfact;\\
&& B_{\rm FOG_5}\stackrel{\text{sq}}{=} \left [ 4\mu_{bk} \mu_k^5 \mu_b - 3\mu_k^4 \mu_b^2 + 2\mu_k^2 \mu_b^4 -
{d\ln P_{\delta}(k)\over d\ln k} \mu_{bk} \mu_k^5 \mu_b \right ]\pfact;\\
&& B_{\rm FOG_6}\stackrel{\text{sq}}{=}
\left [4\mu_{bk} \mu_k^5 \mu_b^3 - \mu_k^4 \mu_b^4 - {d\ln P_{\delta}(k)\over d\ln k}\mu_{bk} \mu_k^5 \mu_b^3 \right]\pfact.
\label{eq:FOGF}
\een

After angular averages of various terms $\la \mu_k^{a}\mu_b^{b}\mu_{bk}^{c}\ra$ are taken and assuming a power-law power spectrum $P(k)\propto k^{n_{\rm eff}}$,
the squeezed bispectrum takes the following form:
\ben
 B_{{\rm X}_i} =\left [{\alpha^{\rm X}_i} -\beta^{\rm X}_i (n_{\rm eff}+3)) \right ]\pfact.
\een
We deduce the following relations:
\ben
\label{eq:SQ1,1}
&& B_{\rm SQ1_1}\stackrel{\text{sq}}{=} {4\over 3}\left [(2+\epsilon)-{1\over 3}(n_{\rm eff}+3)\right ]\pfact \\
&&  B_{\rm SQ1_2} \stackrel{\text{sq}}{=} {8\over 9}\left [(2+\epsilon)-{2\over 9}(n_{\rm eff}+3)\right ]\pfact; \\
&& B_{\rm SQ1_3} \stackrel{\text{sq}}{=}{1\over 225}\left [(72+ 28\epsilon) - {11 \over 225}(n_{\rm eff}+3)\right ]\pfact; \\
&& B_{\rm SQ2_1} \stackrel{\text{sq}}{=}{4\over 9}\left[(2+\epsilon^{\prime}) -{1\over 9}(n_{\rm eff}+3)\right ]\pfact;\\
&& B_{\rm SQ2_2} \stackrel{\text{sq}}{=} {8\over 225}\left [(26+11\epsilon^{\prime}) -{26 \over 225}(n_{\rm eff}+3)\right ]\pfact;\\
&& B_{\rm SQ3_3}  \stackrel{\text{sq}}{=}{4 \over 525}\left [(47 + 9\epsilon^{\prime})- {17 \over 225}(n_{\rm eff}+3)\right ]\pfact. 
\label{eq:SQ2,3}
\een
We have ignored primordial non-Gaussianity in our derivation (see e.g. Ref.\citep{Tellarini} for related discussion).

Notice that, using data from SDSS-III Baryon Osccilation Spectroscopic Survey (BOSS) Data Release 10 CMASS sample, 
a first measurement of position-dependent correlation function has already detected the three-point correlation function
at $7.4\sigma$.  Such measurements are useful in constraining the nonlinear bias of galaxy halos in BOSS CMASS survey \citep{Chinag_thesis}.  

We would like to point out here that the approximations used to simplify a Feldamn-Kaiser-Peacock (FKP)\citep{FKP} type estimator that was used in Ref.\citep{GilMarin_bias}
in their joint analysis of power spectrum and bispectrum fail in the squeezed limit. Thus, the squeezed limit of the bispctrum probed by the position-dependent power spectrum
can provide the useful missing information.

In Figure-\ref{fig:redshift} various contributions to the IB are displayed as a function of the wavenumber $k$. The left panel shows
the contributions from the linear squashing terms of Eq.(\ref{eq:sq1A})-Eq.(\ref{eq:sq1C}), the middle panel corresponds to the second order squashing terms of 
Eq(\ref{eq:sq2A})-Eq.(\ref{eq:sq2C}). Finally, the right panel corresponds to the contribution from Eq.(\ref{eq:FOGA})-Eq.(\ref{eq:FOGF}).
The power spectrum is approximated locally as a power law with a logarithmic slope of $n$ that reduces the 
expressions to the form given in Eq.(\ref{eq:gen}) with coefficients given by Eq.(\ref{eq:SQ1,1})-Eq.(\ref{eq:SQ2,3}).
Results correspond to redshift $z=0$.
\section{Angular (Projected) Integrated Bispectrum}
\label{sec:2D}
Most early studies of galaxy clustering were performed in angular (projected) surveys in 2D.
Due to the presence of huge number of galaxies, a projected survey
allows more precise determination of higher order cumulants and CC.
As an example, using the data from the APM survey, which contains more than $1.3\times 10^6$ galaxies, 
projected cumulants or the $s_n$ parameters were computed up to the ninth order \citep{Gaztanaga} and projected CCs, or the $c_{pq}$ parameters were computed up to
fourth order \citep{SzSz}.
Unlike the redshift surveys the angular or projected surveys do not mix density and velocity
fields which can be difficult to disentangle. The main difficulty with the projects surveys, however,
is that they mix different physical scales. A given angular scale in a projected survey
takes contribution from length that are quasilinear as well as highly
nonlinear length scales \citep{LoVerde}. 
The projected cumulants were derived in Ref.(\citep{APM}), and were later extended to CCs
in Ref.(\citep{WeakStrong}) and Ref.(\citep{BiasLensing}) in the highly nonlinear regime using HA.

In 2D using the kernel defined in Eq.(\ref{eq:F2}) we have the following results:
\ben
\label{eq:den2D}
{B}_{\rm 2D} = \left [(3+\epsilon)-{1\over 2}(n+2)\right ]P_\delta(q_{b\perp})P_\delta(k_{\perp}).
\label{eq:theta2D}
\een
For $\epsilon=3/7$ we recover $\bar{\cal B}^{\delta}(k_{\perp}) = [(24/7)-(n+2)/2]P_{\delta}(q_{q\perp})P_\delta(k_{\perp})$.
See Ref.\citep{MunshiColes} for detailed derivation as well as a survey specific prefactor.
Here $\bk_{\perp}$ is the component of the wavevector orthogonal to the line-of-sight direction, 
i.e., ${\bf k}={\bf k}_{\parallel}+{\bf k}_{\perp};\; k_{\perp}=|{\bf k}_{\perp}|$ and similarly for $\bq_b$.
The corresponding CCs are computed in Ref.(\citep{3Dc21}).
Using the expression for $\epsilon$ from Eq.(\ref{eq:numu}) will allow us to compute the IB in 2D for $\gamma$ models. 
\section{Discussion and Conclusions}
\label{sec:conclu}
Complete characterisation of the bispectrum can be a challenging task as it is a function
of three wavevectors and the shape of the triangle they form. Several techniques 
have been investigated recently in real or Fourier space to reduce the dimensionality
of the problem. In this paper we have studied the well-known CCs in real space and
the position-dependent power spectrum in Fourier domain.

%
%
Generalising the cumulants of biased tracers we introduce the CCs for the biased tracers in \textsection\ref{sec:cc}.
We have combined the halo model predictions and prescriptions from the perturbation theory 
to compute the CCs for collapsed objects Eq.(\ref{eq:J21})-Eq.(\ref{eq:J31}). In doing so, we have computed the
bias parameters $b_k$ from halo models. Combining the cumulants for the collapsed objects 
given in Eq.(\ref{eq:J21})-Eq.(\ref{eq:J31}) with the CCs can be helpful constructing independent
estimates of the bias parameters $b_k$ from simulations and observations.
The particular results that are presented here are for PS and ST mass functions but replacing
these $b_k$ with predictions from other theories can be done in a straight forward manner.  
The results are derived for the collapsed objects but using the $b_k$ for peaks using Eq.(\ref{eq:peak_b1})-Eq.(\ref{eq:peak_b2})
can similarly provides the results for peaks. Our results for the CCs extend the results of Ref.\citep{MoJingWhite} derived for the
cumulants.

%
%
Gravitational clustering induces non-linearity as well as non-locality - bias predicted by the halo model is a local model.
In \textsection\ref{sec:nonlocal} Eq.(\ref{eq:non_local}) we take the phenomenological model proposed in Ref.\citep{McDonaldRoy}.
In addition to parameters $(b_{s^2},b_{s^3},\cdots )$,
this model depends on an infinite set of parameters $( b_{\delta},b_{\delta^2},\cdots )$ 
and makes the density contrast of collapsed objects a function not only of underlying density contrast $\delta(\bx,a)$
but also of the tidal tensor $s$. The amplitudes of these contributions are kept arbitrary.
Combining with generating function approach, in the perturbative regime, we show that the generating function of collapsed 
objects ${\cal G}_h$ can be completely specified only by the generating functions ${\cal G}_\delta$ and ${\cal G}_\theta$
of $\delta(\bx,a)$ and $\theta(\bx,a)$ once the bias parameters are known.

%
%
In \textsection\ref{sec:red} we have considered the IB in the redshift-space.
We have studied the impact of modifying gravity on the redshift-space IB.
We have considered a class of MG theories that were proposed in \citep{Bernardeau_Brax}.
In \textsection\ref{sec:MG} a class of MG theories were introduced. These include DGP brane world
models. In \textsection\ref{sec:red_IB} we investigated the redshift-space IB in a class of MG theories.
The results presented can be generalised to any parametric theories of gravity.

The individual terms that make contributions to the redshift bispectrum are dealt with in \textsection\ref{sec:red_IB}.
These include the linear and non-linear squashing terms. The other terms are the non-linear bias 
terms as well as the Finger-of-God (FOG) terms. The linear and non-linear squashing
terms depend respectively on the bispectrum associated with $\delta$ and $\theta$. 
The squeezed contribution of these terms are derived independently.
These terms are the only terms directly depend on the MG theories.
After angular averaging the IB depends only on the parameters $\epsilon$ and $\epsilon^{\prime}$.
These parameters can be computed using the specific parametrization introduced in \textsection\ref{sec:MG}. 
However, in generic MG theories the bispectrum can take more complicated $(k,a)$ dependence
and the squeezed limits will have to be handled in a case by case manner. Indeed accurate theoretical
modelling of the bispectrum can be demanding even for $\Lambda$CDM and may require many more parameters than
are typically employed \citep{GilMarin}. In \textsection\ref{sec:2D} projected IB is investigated.

Various other forms of data compression for the study of bispectrum have also been developed e.g. skew-spectrum associated with
the gravity induced bispectrum was studied in \citep{PrattenMunshi}. A optimised version of this skew-spectrum
has recently been introduced in \citep{Schmittfull}. Modal estimators were also introduced to study
the gravity induced non-Gaussianity \citep{Regan1,Regan2}. 
Though the IB statistics studied here are sub-optimal it is comparatively simpler to implement and optimisation is possible.
Morphological estimators provide alternate routes to probe non-Gaussianity \citep{PrattenMunshi}.
Minkowski Functionals were studied in redshift-space for the specific model we have studied here \citep{Bernardeau_Minkowski}. 
For a self-consistent way to tackle bias, redshift-space distortion and non-Gaussianity in MG theories an approach similar to 
what has been developed in Ref.\citep{Matsubara1} is required. Combining the generating function approach with such techniques can provide interesting 
clues to statistics of collapsed halos in MG theories.
\section*{Acknowledgements}
\label{acknow}
This work was supported by the Science and Technology
Facilities Council (grant numbers ST/L000652/1). It is a pleasure for the author 
to thank Donough Regan for his help and suggestions to improve the draft and Peter Coles for 
related collaborations. 
\bibliography{paper.bbl}
\appendix
\end{document}